\documentstyle[11pt]{article}
\begin{document}
%%%%%%%%%%%%%%%%%%%%%%%%%%%%%%%%%%%%%%%%%%%%%%%%%%%%%%%%%%%%%%%%%%
\def\sqr#1#2{{\vcenter{\hrule height.4pt\hbox{\vrule width.4pt height#2pt 
\kern#1pt\vrule width.4pt}\hrule height.4pt}}}
\def\Square{\mathchoice{\sqr78\,}{\sqr78\,}\sqr{20.0}{18}\sqr{20.0}{18}} 
%\lta and \gta produce > and < signs with twiddle underneath
\def\spose#1{\hbox to 0pt{#1\hss}}\def\lta{\mathrel{\spose{\lower 3pt\hbox
{$\mathchar"218$}}\raise 2.0pt\hbox{$\mathchar"13C$}}}  \def\gta{\mathrel
{\spose{\lower 3pt\hbox{$\mathchar"218$}}\raise 2.0pt\hbox{$\mathchar"13E$}}} 
\newcommand{\dd}{{\rm d}}
\newcommand{\gsim}{\mathrel{%
   \rlap{\raise 0.511ex \hbox{$>$}}{\lower 0.511ex \hbox{$\sim$}}}}
\newcommand{\lsim}{\mathrel{
   \rlap{\raise 0.511ex \hbox{$<$}}{\lower 0.511ex \hbox{$\sim$}}}}
\newcommand{\ivm}{\overline{\gamma}}					% the projected metric
\newcommand{\norm}{\zeta} 						% the normal to the brane
\newcommand{\tn}{{\cal T}_{\!_\infty}} 					% the bare brane tension
\newcommand{\brT}{\overline{\cal T}} 					% the total brane e-m tensor
\newcommand{\obsT}{\overline{\tau}} 					% the observed part of the brane e-m tensor

%%%%%%%%%%%%%%%%%%%%%%%%%%%%%%%%%%%%%%%%%%%%%%%%%%%%%%%%%%%%%%%%%%
\title{\bf Simulated gravity without true gravity\\
           in asymmetric brane-world scenarios}

\author{Brandon Carter$^1$, Jean--Philippe Uzan$^2$,
        Richard A. Battye$^3$ and Andrew Mennim$^3$}

\vskip0.25cm

\date{(1) DARC/LUTH,  
          Observatoire de Paris, 92195 Meudon Cedex, France, \\ 
      (2) Laboratoire de Physique Th\'{e}orique, UMR 8627 du CNRS, \\  
          Universit\'{e} de Paris XI, B\^atiment 210, 91405 Orsay 
          Cedex, France,\\
      (3) Department of Applied Mathematics and Theoretical Physics,\\
          Centre for Mathematical Sciences,
          University of Cambridge, \\
          Wilberforce Road, Cambridge, CB3 0WA, UK \\ 
\vskip0.25cm
11 June 2001\\}

\maketitle

%%%%%%%%%%%%%%%%%%%%%%%%%%%%%%%%%%%%%%%%%%%%%%%%%%%%%%%%%%%%%%%%%%
\begin{abstract}
This article investigates asymmetric brane-world scenarios in
the limit when the bulk gravity is negligible.  We show that, even
when true self gravity is negligible, local mass concentrations will
be subject to a mutual attraction force which simulates the effect of
Newtonian gravity in the non-relativistic limit. Cosmological
and also post-Newtonian constraints are examined.
\end{abstract}
%%%%%%%%%%%%%%%%%%%%%%%%%%%%%%%%%%%%%%%%%%%%%%%%%%%%%%%%%%%%%%%%%%
\section{Introduction}\label{I}

Gravity has long been considered to be one of four fundamental forces.
Yet it is very different in its nature from the others and, although
particle theories have been devised to unify the other three, it has
proved surprisingly difficult to unify all four.  One possible way to
cut this Gordian knot is to consider gravity not as a fundamental
force but as a derived one~\cite{rio}.  In this paper we consider one
possible way to realise this within the context of higher dimensional
theories by explaining gravity as an apparent effect due to the
acceleration of a brane-world universe through a five dimensional
spacetime.

Following an original suggestion by Randall and Sundrum~\cite{RS99},
many authors have considered 5-dimensional scenarios in which
effectively 4-dimensional (inverse square law) gravity is induced
on a 3-brane hypersurface representing our observed 4-dimensional
spacetime due to the effective confinement of the 5-dimensional gravitational field
by a strong extrinsic curvature. The kind of cosmological scenario
that was originally proposed was characterised by a postulate of Z$_2$
reflection symmetry about the 3-brane hypersurface, in which case the
effective 4-dimensional Newton  constant ${\rm G}_{_{[4]}}$ was found
to be proportional to the square of the analogous 5-dimensional
gravitational coupling constant ${\rm G}_{_{[5]}}$ with a coefficient
proportional to the asymptotic limit value $\tn$ of the brane tension.
A more general relation 
\begin{equation}\label{G4}
{\rm G}_{_{[4]}}={3\over 4\pi\tn}\left(\big(\pi^2{\rm
G}_{_{[5]}}\tn\big)^2 -\Big({\overline f\over 4\tn}\Big)^2\right) \,,
\end{equation} 
has been obtained more recently~\cite{CU01} allowing for the
possible presence of a force density term $\overline f$, such as might
arise naturally from a gauge four form coupling~\cite{BC01}.

Moreover, for scenarios of this more general reflection symmetry
violating kind~\cite{move1,misc,ANO00} it has subsequently been
shown~\cite{Us01} that it is still possible to recover an effective
4-dimensional Einstein equation for local perturbations on the brane
under the same kind of conditions as in the earlier demonstration for
the more specialised reflection symmetric case by Shiromizu {\em et
al}~\cite{SMS00}.  The applicability of the effective 4-dimensional
Einstein equations~\cite{Us01,SMS00} with the coupling constant given
by (\ref{G4}) is restricted to the weak source limit (relevant for the
present epoch but not for earlier stages of cosmological evolution
during which gravity would have behaved in a ``non-conventional''
manner~\cite{BDL00}) and to the case where the background Weyl tensor
contributions, representing the effect of incoming gravitational waves
or black holes in the five dimensional ``bulk'', are sufficiently
small.

The purpose of the present work is to consider a very different kind
of limit, involving a brane-world scenario in which an inverse
square law for gravity is recovered, not by an effective confinement
mechanism due to strong extrinsic curvature, but simply by having a
sufficiently small five dimensional gravitational coupling constant
for genuine gravitation to be entirely negligible.  In this weak
gravity limit our previous formula (\ref{G4}) will cease to be
applicable otherwise an, unacceptable, negative value for ${\rm
G}_{_{[4]}}$ would be obtained: in the limit considered here, the
derivation of (\ref{G4}) for localised perturbations~\cite{Us01} will
break down because the Weyl contribution (on whose
negligibility its validity depended) will become indeterminate due to
${\rm G}_{_{[5]}}$ terms in the denominator.  In order to deal with
this limit, which is singular from the previous point of view, we need
to start again with a fresh approach.
 
The preceding work~\cite{Us01} shows that the worldsheet geometry,
and hence the apparent gravity on the brane, has three origins.  There
is a first part coming from the geometry of the ``bulk'' background, a
second part arising from the discontinuity of the extrinsic curvature
form, $K_{\mu\nu}$, across the worldsheet, and a third part coming
from its mean value. In the reflection symmetric case that is most
commonly considered, the latter effect is absent (that is, there is a
vanishing mean $\langle K_{\mu\nu}\rangle=0$). The aim of the present
article is to investigate the opposite extreme limiting case
characterised by the absence of gravitational perturbations and, in
particular, of the gravitationally engendered discontinuity of the
extrinsic curvature (that is, we will have $[K_{\mu\nu}]=0$) so
that it is the mean curvature, identifiable in this case with the
value on either side, which provides the dominant contribution.

What will be shown here is that, when five dimensional gravity is
absent or negligible, an artificial, and automatically positive, 
4-dimensional (inverse square law) gravitation effect will still be
present whenever reflection symmetry is broken, with an effective
coupling ``constant'' that will be proportional to the rest frame time
component, $K_{_{00}}$ say, of the extrinsic curvature of the averaged
background, according to a formula which, instead of (\ref{G4}) will
have the form)
\begin{equation}\label{G4a} {\rm G}_{_{[4]}}={K_{_{00}}^{\,2}\over 
4\pi\tn}\, .\end{equation} However although it mimics Newtonian
gravitation in the non-relativistic limit, the artificial gravitation
occurring in the simple type of 5-dimensional 3-brane scenario
considered here will produce post-Newtonian effects that will, in
general, differ from those of the standard Einstein theory, typically
in a parameter range beyond what is observationally admissible.

Although it will be shown that it can be contrived in such a way as to
satisfy the cosmological nucleosynthesis constraint and
instantaneously mimicking the standard Einstein theory in the
immediate spacetime neighbourhood of our solar system, the necessary
fine tuning would seem to be very artificial.  Furthermore, a
gravitational ``constant'' that is proportional to the, possibly
variable, extrinsic curvature of the average background can hardly be
compatible with the very severe observational limits on deviations
from constancy of gravitational coupling.

These considerations lead us to conclude that in its present form the
simplest model considered here does not by itself provide a viable
alternative theory of gravity. However --- quite apart from the
possibility that a viable alternative may be provided by some
generalisation --- the simulated worldsheet gravity
phenomenon considered here will still be of physical interest and
potential practical relevance as a mechanism that can provide small
but not necessarily negligible additions to effects of genuinely
gravitational origin (in much the same way as, from a laboratory point
of view, the genuine gravitational attraction of our Earth is modified
by the centrifugal effect of its rotation).

The plan of this article is as follows. The necessary worldsheet
perturbation formalism for the general case of a codimension one brane
of arbitrary worldsheet dimension $p$ is developed in section~\ref{II}.
The relevant non-gravitating kind of minimally
coupled~\cite{BC01} homogeneous isotropic cosmological background to
which it will be applied is presented next in section~\ref{III}, where
the first integrated (Friedmann type) form of the dynamical equations
is obtained in terms of an adjustable constant of integration (whose
interpretation as the energy with respect to the stationary geometry
of the external ``bulk'' is described in the appendix). The heart of
this article is in section~\ref{IV} where it is shown how local
perturbations will generically be subject to an attraction of
Newtonian gravitational in the non-relativistic limit, and where the
more restrictive conditions for agreement with the linearised Einstein
equations are derived.  The remainder of the discussion is restricted
to the cosmologically interesting case of a hypermembrane with
worldsheet dimension $p=4$, starting in section~\ref{V} where the
conditions for compatibility with the standard description of
nucleosynthesis are derived. In conclusion, the difficulty of
reconciling these global conditions with what is required for matching
the parameters governing local gravitational effects is discussed in
section~\ref{VI}.

%%%%%%%%%%%%%%%%%%%%%%%%%%%%%%%%%%%%%%%%%%%%%%%%%%%%%%%%%%%%%%%%%%%%%%
\section{Lagrangian worldsheet perturbations}\label{II}

As a technical prerequisite for the perturbation analysis that will be
involved, we start by considering the effect of the brane worldsheet
perturbation, which we suppose to be generated by an infinitesimal
displacement vector field $\xi^\mu$ say, on the first and second
fundamental forms of the worldsheet, which correspond to tensors
denoted by $\ivm_{\mu\nu}$ and $K_{\mu\nu}$ using the
notation of our previous work~\cite{Us01}. In terms of the background
spacetime metric $g_{\mu\nu}$ and the unit normal
$\norm_\mu=\nabla_{\!\mu}\zeta$, where $\zeta$ measures orthogonal
distance from the brane (on the side chosen to be considered as
positive) the first fundamental tensor is defined simply by
\begin{equation}\label{06} 
\ivm_{\mu\nu}\equiv g_{\mu\nu} - \perp_{\mu\nu}\, ,
\qquad
\perp_{\mu\nu}\equiv\norm_\mu\norm_\nu \, .
\end{equation}
In terms of the tangentially projected differentiation operator
\begin{equation} 
\overline\nabla_{\,\mu}\equiv\ivm_\mu^{\ \nu}\nabla_{\!\nu}
\end{equation} 
the tensor corresponding to the second fundamental form is
given\footnote{We are using the MTW sign conventions~\cite{MTW}.} by
\begin{equation}\label{08} 
K_{\mu\nu}=-\overline\nabla_{\!\mu}\norm_\nu \, ,
\end{equation}
and will automatically have the Weingarten symmetry property
$K_{\mu\nu}=K_{\nu\mu}$ as a worldsheet integrability condition.
This tensor is also obtainable from the general second fundamental
tensor $K_{\mu\nu}{^\rho}$ that was used in an earlier brane
perturbation analysis~\cite{BC95} by $K_{\mu\nu} = K_{\mu\nu}{^\rho}\norm_\rho$.

Since our purpose here is to consider the limit in which genuine
gravitation is negligible, the geometry will be globally unaffected by
the perturbation, so the Eulerian metric perturbation $h_{\mu\nu}$ $=
\delta_{_{\rm E}} g_{\mu\nu}$ that was allowed for in the earlier
analysis~\cite{BC95} can simply be set to zero.  Thus locally, with
respect to coordinates that are comoving with respect to the
displacement (so that the brane worldsheet retains the same
coordinate locus, $\zeta=0$) the only change in the metric tensor will
be the purely Lagrangian variation given by the familiar Lie
differentiation formula
\begin{equation}\label{10} 
\delta_{_{\rm L}} g_{\mu\nu}=2\nabla_{(\mu}\xi_{\nu)}\, ,
\end{equation}
using round brackets to indicate index symmetrisation. Preservation of
the locus of the worldsheet implies preservation of the direction of
the covariant normal $\norm_\mu$, but the change (\ref{10}) in the
metric will entail a corresponding change in amplitude to preserve the
unit normalisation property $\norm_\mu\norm^\mu=1$, so it can be seen
that the ensuing Lagrangian variation of the unit normal will be given
by
\begin{equation}\label{11} 
\delta_{_{\rm L}}\norm_\mu=\norm_\mu\norm_\nu\norm_\rho
\nabla^\nu\xi^\rho\, .
\end{equation}
This formula can be used to obtain a direct verification of the
corresponding Lagrangian variation formula for the first fundamental 
tensor, namely
\begin{equation}\label{12}  
\delta_{_{\rm L}}\ivm^{\mu\nu}=
-2\ivm_\sigma^{\ (\mu} \overline\nabla{^{\nu)}}\xi^\sigma\, ,
\end{equation}
which is actually valid, not just for the hypermembrane (codimension
1) case under consideration here, but quite generally for a
hyperstring (the case with codimension 2, as in a 6-dimensional world
brane scenario~\cite{GSh00}) and cases with even higher codimension.

Using the explicit expression
\begin{equation}\label{14} 
\nabla_{\!\mu}\norm^\nu=\norm^\nu_{\, ,\mu}+
\Gamma^{\ \nu}_{\!\mu\, \rho}\, \norm^\rho\,, \qquad
\Gamma^{\ \nu}_{\!\mu\, \rho}= g^{\nu\sigma}\left(g_{\sigma(\mu , \rho)}
-{_1\over^2}g_{\mu\rho ,\sigma}\right) \, ,
\end{equation}
for the covariant derivative that is involved in (\ref{08})
and the well known  fact that the Lie derivative
\begin{equation}\label{15} 
\delta_{_{\rm L}}\Gamma^{\ \nu}_{\!\mu\, \rho}= 
g^{\nu\sigma}\left(\nabla_{\!(\rho}\delta_{_{\rm L}}g_{\mu)\sigma}
-{_1\over^2}\nabla_{\!\sigma}\delta_{_{\rm L}}g_{\mu\rho}\right) \, ,
\end{equation}
of the Christoffel connection can be expressed in terms of the 
background Riemann tensor ${\cal R}_{\mu\nu\rho\sigma}$ as
\begin{equation}\label{16} 
\delta_{_{\rm L}}\Gamma^{\ \nu}_{\!\mu\, \rho}=\
\nabla_{\!(\mu} \nabla_{\rho)}\xi^\nu-{\cal R}^\nu_{\ (\mu\rho)\sigma}
\xi^\sigma\, .
\end{equation}
Together with (\ref{11}) and (\ref{12}), this enables us to calculate
the corresponding variation of the extrinsic curvature as
\begin{equation}\label{18}
\delta_{_{\rm L}} K^{\mu\nu}=\norm_\rho
\ivm^{(\mu}{_\kappa}\ivm^{\nu)}{_\lambda}
\left(\overline\nabla{^{(\kappa}} \overline\nabla{^{\lambda)}}
\xi^\rho +{\cal R}^{\kappa\rho\lambda}{_\sigma}
\xi^\sigma\right)+4K_\sigma^{\,(\mu}\overline\nabla{^{\nu)}}
\xi^\sigma\, .
\end{equation}
In the language of~\cite{BC95}, this can be obtained from the
formula $K_{\mu\nu} $ $=K_{\mu\nu}{^\rho}\norm_\rho$ using (\ref{11})
and the general expression for $\delta_{_{\rm L}} K_{\mu\nu}{^\rho}$.
If one makes the convenient choice of an orthogonal gauge, setting
\begin{equation}\label{20} 
\xi^\mu=\xi\, \norm^\mu\, ,
\end{equation}
it can be seen that (\ref{18}) will be expressible directly
in terms of the scalar displacement amplitude $\xi$ as
\begin{equation}\label{21}
\delta_{_{\rm L}} K^{\mu\nu}=
\ivm^{(\mu}{_\kappa}\ivm^{\nu)}{_{\!\lambda}}
\left(\overline\nabla{^{(\kappa}} \overline\nabla{^{\lambda)}}\xi 
+{\cal R}^{\kappa\rho\lambda}{_\sigma}\!\perp^{\!\sigma}_{\,\rho}
\xi\right)+3K^{\sigma\mu}K_{\!\sigma}^{\,\nu}
\xi\, .
\end{equation}
In what follows we shall be particularly interested in the trace
$K=K_{\!\nu}^{\,\nu}$, for which the corresponding variation can be
seen to be expressible in terms of the background Ricci tensor,
${\cal R}_{\mu\nu}={\cal R}_{\sigma\mu}{^\sigma}{_\nu}$, as
\begin{equation}\label{25} 
\delta_{_{\rm L}} K=\overline\nabla{^\mu}
\overline\nabla_{\!\mu}\,\xi +\left(K^{\mu\nu}K_{\mu\nu}+
\perp^{\!\mu\nu}\!{\cal R}_{\mu\nu}\right)\xi \, .
\end{equation}

This formula leads to a noteworthy cancellation when used in
conjunction with the contraction of the well known Gauss identity (see,
for example, the preceding analysis~\cite{C00}) according to which the
Ricci scalar $\overline R$ of the induced metric on the hypersurface
will be given by
\begin{equation}\label{30}
\overline R=K^2-K^{\mu\nu}K_{\mu\nu}  +
2\ivm^{\mu\nu}{\cal R}_{\mu\nu}-{\cal R}
\, .
\end{equation}
The preceding formulae are valid regardless of the dimension, $p$  say, 
of the worldsheet under consideration, the only restriction being 
that it should have codimension one.  However 
we must take account of this dimension in the next step, which depends 
on separating the trace-free part of the extrinsic curvature tensor,
\begin{equation}\label{32} 
C_{\mu\nu}=K_{\mu\nu}-{1\over p}K \,
\ivm_{\mu\nu}\,, \qquad C_\nu^{\,\nu}=0\, .
\end{equation}
and also the trace-free part of the background Ricci tensor,
\begin{equation}\label{33} 
{\cal S}_{\mu\nu}={\cal R}_{\mu\nu}-{1\over p+1}{\cal R}
g_{\mu\nu}\,, \qquad {\cal S}_\nu^{\,\nu}=0\, ,
\end{equation}
which vanishes whenever the background satisfies Einstein type vacuum
equations, with or without a cosmological constant.  What is
remarkable is that the form of (\ref{25}) is such that the identity
(\ref{30}) can be used to eliminate not just the extrinsic curvature
trace $K$, but also the background Ricci trace scalar, ${\cal R}$, in
favour just of the intrinsic Ricci trace scalar $\overline R$, and hence
\begin{equation}\label{35} 
\delta_{_{\rm L}} K=\overline\nabla{^\mu}
\overline\nabla_{\!\mu}\,\xi + {\xi\over p-1}\left(\overline R+ p\,
C^{\mu\nu} C_{\mu\nu} -({p+1})\overline
\gamma^{\mu\nu}{\cal S}_{\mu\nu}
\right)\, .
\end{equation}
This generalises the results by Garriga and Vilenkin~\cite{gv}.  The
particular case $p=2$ of this identity applies to any ordinary
2--surface in 3-dimensional space, while the case $p=3$ applies to the
worldsheet of an ordinary membrane, including the case of a
cosmological domain boundary, in 4-dimensional spacetime. What we
shall be concerned with here is the case $p=4$ of a 3-brane in a
5-dimensional spacetime background. In this latter context, the
significant feature of (\ref{35}) is that in typical brane-world
scenarios the quantities in the bracket on the right are all zero or
tend to zero at late cosmological times, unlike the cancelled traces
$K$ and ${\cal R}$, whose amplitudes remain very large.

%%%%%%%%%%%%%%%%%%%%%%%%%%%%%%%%%%%%%%%%%%%%%%%%%%%%%%%%%%%%%%%%%
\section{Homogeneous cosmological reference configuration}
\label{III}
 
The previous analysis will allow us to study the Newtonian limit of
the theory. First, we shall study homogeneous and isotropic reference
configuration which will also be of interest for the cosmological
solutions.  Our purpose now is thus to consider the case of a
hypermembrane governed by an equation of motion taking the standard form
\begin{equation}\label{40} 
\brT{^{\mu\nu}}K_{\mu\nu}=\overline f\, ,
\end{equation}
that applies~\cite{C00} in the limit when self gravitation is
negligible so that the background geometry can be considered to be
smooth. In this dynamical equation, $\overline f$ is the orthogonal
component of the external force density, and $\overline {\cal
T}{^{\mu\nu}}$ is the surface stress energy density tensor, which will
satisfy an intrinsic pseudo--conservation law of the form
\begin{equation}\label{42} 
\ivm_{\nu\rho}\overline\nabla_{\!\mu}
\brT^{\mu\nu}=0\, ,
\end{equation}
provided the external force density has no tangential component.  We
shall restrict our attention to the kind of force produced by minimal
gauge coupling~\cite{CU01,BC01} for which this orthogonality condition
will automatically be satisfied, and for which the force
density $\overline f$ will simply be a constant.

We wish to study perturbations of a homogeneous isotropic reference
configuration in a background satisfying Einstein type vacuum
equations of the standard form
\begin{equation}\label{46} 
{\cal S}_{\mu\nu}=0 \, .
\end{equation}
As discussed in the preceding work~\cite{CU01}, application of the
relevant generalisation~\cite{BCG} of the well known 4-dimensional
Birkhoff theorem~\cite{MTW} allows us to conclude that the background
metric will have the static form
\begin{equation}\label{50} 
\dd s^2= r^2 \dd\ell^2 +{\dd r^2\over{\cal V}}-{\cal V} \dd t^2
\, ,
\end{equation}
where $\dd\ell^2$ is the positive-definite space metric of a
$(p-1)$-dimensional sphere, plane, or anti-sphere with constant
curvature, $k$ say, respectively positive, zero, or negative, and
${\cal V}$ is a function only of $r$. The location of the brane
worldsheet will be given by an expression of the form $r=a$ where $a$
is a function only of $t$. There will be a equivalent brane based
formulation
\begin{equation}\label{51} 
\dd s^2= r^2\, \dd\ell^2+\dd\zeta^2 -\nu^2 \dd\tau^2\,
\end{equation}
in which the worldsheet locus is given simply by $\zeta=0$, and in
which $\tau$ represents proper time on this locus, so that $\nu=1$
there. Using a dot to denote partial differentiation with respect to $\tau$,
and a dash for partial differentiation with respect to $\zeta$, the
rate of change of the scale factor $a$ will satisfy the conditions
\begin{equation}\label{52} 
\dot a ^2=r^{\prime 2}-{\cal V}\, \qquad
{\rm and}\qquad
\dot a ^2= \dot r^2/\nu^2\, .
\end{equation}
In such a system, the first fundamental form (\ref{06})
specifying the induced metric on the worldsheet, is
\begin{equation}\label{54} 
\dd\overline s^2=\ivm_{\mu\nu}\, \dd x^\mu\, \dd x^\nu=
a^2\, \dd\ell^2-\dd\tau^2\, ,
\end{equation}
and the corresponding expression for second fundamental form
(\ref{08}) is
\begin{equation}\label{55} 
K_{\mu\nu}\,\dd x^\mu\,\dd x^\nu =-a r^\prime \, \dd\ell^2
+\nu^\prime\, \dd\tau^2\, ,
\end{equation}
which can also be expressed as
\begin{equation}\label{56} 
K_{\mu\nu}= -{r^\prime\over a}\ivm_{\mu\nu}
+\left(\nu^\prime -{r^\prime\over a}\right) \overline u_\mu\overline u_\nu\, .
\end{equation}
in terms of the preferred timelike unit vector $\overline
u_\nu\equiv\nabla_{\!\nu} \tau$ that specifies the cosmological reference
frame. In terms of this unit vector, the stress-energy momentum tensor
will have the isotropic perfect fluid form
\begin{equation}\label{62} 
\brT^{\mu\nu}= -{\cal T}\,\ivm^{\mu\nu}
+({\cal U}-{\cal T})\overline u^\mu \overline u^\nu\, ,
\end{equation}
where ${\cal U}$ is the total rest frame energy density and ${\cal T}$
is the total brane tension.  It can be deduced from (\ref{55})
that the dynamical equation (\ref{40}) takes the form
\begin{equation}\label{66} 
{\cal U}\nu^\prime+({p-1}){\cal T} {r^\prime\over a}
=\overline f\, 
\end{equation}
from which we deduce that the extrinsic curvature scalar is given by
\begin{equation}\label{57}  
K=-{\overline f\over{\cal U}}-({p-1})\left(1-
{{\cal T}\over{\cal U}}\right){r^\prime\over a}\, ,
\end{equation}
while the trace-free part will have the form
\begin{equation}\label{58} 
C_{\mu\nu}=C\,\left({1\over p}\ivm_{\mu\nu}
+\overline u_\mu \overline u_\nu\right)\, ,
\end{equation}
so that the scalar in (\ref{35}) is explicitly given by
\begin{equation}\label{59}  
C^2= {p\over p-1}C^{\mu\nu}C_{\mu\nu}=
\left(\nu^\prime -{r^\prime\over a}\right)^2\, .
\end{equation}

Due to homogeneity, the non-trivial content of the
pseudo-conservation law (\ref{42}) reduces to the simple energy
equation
\begin{equation}\label{64} 
\dot{\cal U}+({p-1})\left({\cal U}-{\cal T}\right)
{\dot a\over a}=0\, .
\end{equation}
Using (\ref{64}) and the relation $\dot r^\prime=\nu^\prime\dot a$
obtained from (\ref{52}), the dynamical equation (\ref{66}) can be
rewritten as
\begin{equation}\label{68} 
\frac{\dd}{\dd\tau}\left({\cal U} r^\prime a^{p-1}\right)
={\overline f\over p}\frac{\dd}{\dd\tau}\left(a^p\right)\, .
\end{equation}
Since, as remarked above, the minimal coupling postulate
automatically ensures~\cite{CU01,BC01} that $\overline f$ is constant,
this equation will be immediately integrable to give
\begin{equation}\label{70} 
r^\prime={a\over{\cal U}}\left({\overline f\over p}+
{E_{_{\rm B}}\over a^p}\right) \, ,
\end{equation}
where $E_{_{\rm B}}$ is a constant of integration, which can be
interpreted as being proportional to the global energy of the system. 
Since the concept of energy normally involves reference to some stationary
background reference system so as to be well defined, it can not be
used in the usual global cosmological context of an expanding universe 
model. It will, however, be shown in the appendix how a bulk
background which is static --- due to the generalised Birkhoff
theorem~\cite{BCG} --- will allow us to define this energy in an
unambiguous manner for minimally coupled scenarios of the kind
considered here. Using (\ref{66}) again, we obtain the corresponding
formula 
\begin{equation}\label{71} 
\nu^\prime = {\overline f\over{\cal U}}-({p -1})
{ {\cal T}\over {\cal U}^2} \left({\overline f\over p}+
{E_{_{\rm B}}\over a^p}\right) \, ,
\end{equation}
which completes the evaluation of the coefficients in the expression
(\ref{55}) for the second fundamental form. Therefore, the extrinsic
curvature scalar is given by
\begin{equation}\label{72} 
K=-{\overline f\over  p\,{\cal U}}\left( 1+({p-1})
{{\cal T}\over{\cal U}}\right)-{{p-1}\over{\cal U}}\left(
1-{{\cal T}\over{\cal U}}\right){E_{_{\rm B}}\over a^p}\, ,
\end{equation}
and the corresponding expression for the scalar coefficient $C$ in
(\ref{58}) and (\ref{59}) is
\begin{equation}\label{73} 
C={({p-1})\overline f\over p\,{\cal U}}\left(
1-{{\cal T}\over \rm {\cal U}}\right) -{1\over{\cal U}}\left(1+({p-1})
{{\cal T\over\cal U}}\right){E_{_{\rm B}}\over a^p}\, .
\end{equation}

As in the preceding work~\cite{CU01,BC01} on the self gravitating case, 
we adopt the usual supposition that the hypersurface stress energy 
tensor $\brT{^{\mu\nu}}$ in (\ref{40}) has the form
\begin{equation}\label{75}  
\brT{^{\mu\nu}}=\tn \ivm^{\mu\nu}+\obsT^{\mu\nu}\, ,
\end{equation}
in which $\obsT^{\mu\nu}$ is the part attributable to matter fields 
on the brane and $\tn$ is a constant interpretable as the 
``bare'' brane tension meaning the limit to which the tension and energy 
density tend in the infinitely distended limit. In the homogeneous case 
characterised by (\ref{62}) the material contribution will be given in 
terms of average cosmological density and pressure functions
\begin{equation}\label{76} 
\overline\rho={\cal U}-\tn\, ,\qquad
\overline P=\tn-{\cal T}\, ,
\end{equation}
by
\begin{equation}\label{77} 
\obsT^{\mu\nu}= \overline P\,\ivm^{\mu\nu}
+(\overline\rho+\overline P)\overline u^\mu \overline u^\nu\, .
\end{equation}

In the large expansion limit $a\rightarrow\infty$ that is relevant
at late cosmological times the material contributions will
tend to zero, $\overline\rho\rightarrow 0$ and $\overline P\rightarrow 0$,
so that ${\cal U}\rightarrow \tn$ and ${\cal T} \rightarrow \tn$,
which implies that we shall have 
\begin{equation}\label{78}
K\rightarrow K_{_\infty}\equiv-{\overline f\over \tn}
\,, \qquad C\rightarrow 0\,, 
\end{equation}
which will be discussed in more details in the following sections.

\section{Perturbed Configuration}\label{IV}

We now turn our attention to an ``actual'' configuration in which the 
brane worldsheet deviates from that of the homogeneous reference 
configuration described above by a small displacement with amplitude 
$\xi$ of the kind described in section~\ref{II}. Using a tilde to 
distinguish perturbed quantities from their analogues in the reference 
configuration, we take the  first and second 
fundamental tensors of the perturbed worldsheet to be given, using the 
notation of (\ref{16}) and (\ref{18}), by
\begin{equation}\label{100}
\widetilde{\ivm}_{\mu\nu} =\ivm_{\mu\nu}
+\delta_{_{\rm L}}\ivm_{\mu\nu}\, ,
\end{equation}
and
\begin{equation}\label{101}\widetilde K_{\mu\nu} =K_{\mu\nu}
+\delta_{_{\rm L}} K_{\mu\nu}\, .
\end{equation}
The corresponding perturbed version of the equation of motion (\ref{40}) 
will have the form
\begin{equation}\label{103} 
\widetilde{\brT}{^{\mu\nu}}\widetilde K_{\mu\nu}=\overline f\, ,
\end{equation}
with a perturbed surface energy-momentum tensor of the form
\begin{equation}\label{104}
\widetilde{\brT}{^{\mu\nu}}=-\tn
\widetilde{\ivm}{^{\mu\nu}}+\widetilde{\obsT}{^{\mu\nu}}\, ,
\end{equation}
in which the constants $\tn$ and $\overline f$ are just the 
same as in the reference configuration. 

By subtracting (\ref{103}) from its unperturbed analogue (\ref{40}) we
are left with a dynamical source equation for $\xi$ having the form
\begin{equation}\label{108} 
\delta_{_{\rm L}} K=K_{\mu\nu}\left(\widetilde
\epsilon{^{\mu\nu}}-\epsilon^{\mu\nu}\right)+
\epsilon^{\mu\nu}\delta_{_{\rm L}} K_{\mu\nu}\, ,
\end{equation}
with $\delta_{_{\rm L}}K$ as given by (\ref{35}) and where we have introduced  
the recalibrated dimensionless material stress-energy contribution
\begin{equation}\label{106} 
\epsilon^{\mu\nu}\equiv{1\over \tn}\,
\obsT^{\mu\nu}\, ,
\end{equation}
When $\epsilon_{\mu\nu}=0$ and $\widetilde\epsilon_{\mu\nu}=0$,
equation (\ref{108}) can be interpreted, using (\ref{35}), as the
equation of evolution for a scalar field $\xi$ non-minimally coupled
to the Ricci scalar, $\overline R$, and with an effective mass given
(of the kind discussed by Garriga and Vilenkin~\cite{gv}) by
\begin{equation}
m^2_{_\xi}=-\frac{1}{p-1}\left(p\,
C^{\mu\nu} C_{\mu\nu} -({p+1})\overline
\gamma^{\mu\nu}{\cal S}_{\mu\nu}\right).
\end{equation}

\subsection{Simulation of Newtonian gravity}

Let us now consider the field in the neighbourhood of a concentrated
non-relativistic source, that is one whose energy-momentum
contribution is given in terms of a preferred timelike unit reference
vector $\overline u^\mu$ by the approximate formula
\begin{equation}\label{110} 
\widetilde{\obsT}^{\mu\nu}-\bar\tau^{\mu\nu}\simeq 
\delta\rho\,\overline u^\mu \overline u^\nu \, .
\end{equation}
We consider a configuration that deviates from an almost uniform and
low curvature reference configuration by the presence of a matter
distribution within a region which is small compared to the reference
curvature scale, so that we can work at first order both in
$\epsilon_{\mu\nu}$ and in the displacement $\xi$.  Hence, the second
term of the right hand side of (\ref{108}) is of order
${\cal O}(\epsilon\xi)$ and will thus be negligible compared with the
first term. On short enough lengthscales, we can 
keep only the gradient terms of highest order, so that we have
that $\delta_{_{\rm L}}K\simeq\overline{\,\Square}\xi$ and the
preceding source equation for $\xi$ will reduce to the simple form
\begin{equation}\label{111}
\overline{\,\Square}
\xi\simeq {K_{_{00}}\over \tn}\,
\delta\rho \,,
\end{equation}
using the notation
\begin{equation}\label{112} 
\overline{\,\Square}\equiv\overline\nabla{^\mu}\overline
\nabla_{\!\mu}\, ,
\qquad K_{_{00}}=K_{\mu\nu}\overline u^\mu\overline u^\nu \, .
\end{equation}
This approximation is the same as the one used when one 
neglects the cosmological fluids with respect to the mass of the
Sun on length scales of order of the Solar system.

The operator $\overline{\,\Square}$ is interpretable as the wave operator
of the local spacetime geometry on the brane,  and $K_{00}$ is the 
time component of the extrinsic curvature with respect to the preferred 
rest frame, which is interpretable simply as the component of its
acceleration orthogonal to the worldsheet.  In terms of
$p$-dimensional worldsheet coordinates 
$\sigma^i$, with respect to which the induced metric has components 
$\gamma_{ij}$ and determinant $\vert\vert\gamma\vert\vert$,  
(\ref{112}) can be replaced by the more explicit expressions 
\begin{equation}\label{112a} 
\overline{\,\Square}\xi =\Vert\gamma\Vert^{-1/2}\left(
\Vert\gamma\Vert^{1/2}\gamma^{ij}\xi_{, i}\right)_{ , j}\, ,
\qquad K_{_{00}}=K_{ij} \overline u^i  \overline u^j \, .
\end{equation}

Using (\ref{08}), (\ref{12}) and (\ref{20}), the resulting variation
of the first fundamental tensor $\ivm_{\mu\nu}$ can be seen to be
\begin{equation} 
\delta_{_{\rm L}}\label{113a}     \ivm_{\mu\nu}=-2 K_{\mu\nu}\,\xi\, .
\end{equation}
Since, on the local scale under consideration, the unperturbed intrinsic 
metric can be taken to be of flat Minkowski form, $\gamma_{ij}=\eta_{ij}$
say, the corresponding perturbed form of the induced metric will be 
expressible in the form
\begin{equation}\label{113} 
\widetilde \gamma_{ij}=\eta_{ij} + h_{ij}\, ,
\end{equation}
in which, by (\ref{113a}), the induced metric perturbation will be given by
\begin{equation}\label{114}
h_{ij}=-2 K_{ij}\,\xi \, .
\end{equation}

For a spherical or hyperspherical static source distribution with
integrated mass $\widetilde M$, the equation (\ref{111}) for $\xi$ has
the well known solution
\begin{equation}\label{115}
\xi =\left({-K_{_{00}}\over{(p-3)}\Omega^{[p-2]}
\,\tn}\right) {\widetilde M\over \overline r^{\,p-3}}
\,,
\end{equation}
for $p\geq4$, while for the case of a circularly symmetric source
distribution on a membrane in an ordinary 4-dimensional spacetime
background the solution is
\begin{equation}\label{115a}
\xi=\frac{K_{_{00}}}{2\pi \tn}\widetilde{M}\ln\,\overline r \,,
\end{equation}
where, in both cases, $\overline r$ is the radial distance from the centre.

The rest frame component
\begin{equation}\label{116}  
h_{_{00}}=h_{ij}u^i u^j\, ,
\end{equation}
of this perturbation is given in terms of the corresponding rest
frame component $K_{00}$ of the second fundamental form by the
dimensionally generalised Newtonian limit  formula 
\begin{equation}\label{117} 
h_{_{00}} =2{\rm G}_{[p]}\, {\widetilde M\over \overline r^{\,p-3}}\, ,
\end{equation}
if one identifies the unrationalised $p$-dimensional gravitational 
constant as
\begin{equation}\label{118} 
{\rm G}_{[p]} = {K_{_{00}}^{\,2}\over{(p-3)}\Omega^{[p-2]}
\, \tn} \qquad(p>3)\, .
\end{equation}

We now concentrate on the constraints that have to be taken into account
in the case $p=4$ which is of greatest interest to us but we will also
we then discuss some applications of the $p=3$ case.

\subsection{Observational constraints when $p=4$}

It seems from (\ref{116}) and (\ref{118}) that this mechanism can
provide a plausible theory for Newtonian gravitation but there are
some important additional constraints that have to be taken into account.
In this section we shall start by considering the constraint arising from the
requirement that as well as reproducing Newtonian effects, the theory
should also have the correct relativistic behaviour in the weak-field
limit, meaning that it should for instance agree with solar system
measurements of gravity.

With respect to an unperturbed metric of the standard Minkowski form,
$\eta_{_{00}}= -1$, $\eta_{{0a}}=0$, $\eta_{ab}=\delta_{ab}$ (using Latin 
space indices, $a=1,2,3$) a perturbation $h_{ij}$ of the form introduced
in (\ref{113}) will be expressible in terms of the relevant parameters of 
standard post-Newtonian parameterisation~\cite{will} by the formulae 
\begin{eqnarray}\label{PPN}
h_{_{00}}&\simeq&2U-2\beta\, U^2\, , \nonumber\\
h_{0a}&\simeq&-\frac{1}{2}(\alpha_1-2\alpha_2)
                      w_a U\, ,\nonumber\\
h_{ab}&\simeq&2\gamma \,U\delta_{ab}\,
\end{eqnarray}
so that the trace $h=h^i_{\, i}$ will be given by the formula
\begin{equation}\label{PPNa} h=2(3\gamma-1)U+2\beta\, U^2\, ,\end{equation}
in which the quantities $w_a$ are the velocity components of the PPN
coordinate system relative to the mean rest-frame of the universe
(which is usually interpreted to be that of the cosmic microwave
background) and $U$ is the Newtonian potential, whose asymptotic form
at large distances from the source will of course be given by $U \sim
{\rm G}_{_{[4]}}\tilde M/\bar r$. The standard post-Newtonian
parameters $\gamma$, $\beta$, $\alpha_1$ and $\alpha_2$ are fixed
physical constants.

Any gravitational theory should predict values for these constants
and confront the results with what is observed.  In particular,
Einstein's general relativity gives that $\gamma=1$, $\beta=1$ and that all
other post-Newtonian parameters must be zero. Less restrictively,
Jordan - Brans - Dicke theory involves a free parameter $\omega$ in terms
of which $\gamma=(\omega+1)/(\omega+2)$, while giving $\beta=1$ and
zero values for the other post-Newtonian parameters~\cite{bd}. 
The freedom to adjust the parameter $\omega$ can be
used~\cite{br} to match the the Jordan-Brans-Dicke theory
to a generic configuration of the form (\ref{114}). However this freedom 
ceases to exist when measurements of gravitational light deflection,
and the Shapiro time delay effect, are taken into account. These
effects provide an observational bound on the PPN parameter $\gamma$ 
whose current value is given by
\begin{equation} \label{obsgam}
\left|\gamma-1 \right|<3\times 10^{-4}.
\end{equation}
This quantity measures how much space-curvature is produced by a unit rest 
mass, whereas the quantity $\beta$ measures the degree of ``non-linearity'' in 
the superposition law of gravity, an effect whose estimation is beyond the
scope of the purely linear analysis  provided here. By substituting (\ref{114}) 
in (\ref{PPN}) and (\ref{PPNa}) it can be seen that what our linear analysis 
does provide is a limit
\begin{equation}\label{123a}
\left|\frac{K}{K_{_{00}}}-2\right|<9\times10^{-4}
\end{equation}
that must somewhow be satisfied by the extrinsic curvature as a condition for 
compatibility with the observational limit (\ref{obsgam}) on the assumption 
that the unperturbed configuration is spatially isotropic. 

Apparent deviations from isotropy might arise from the effects of the 
parameters $\alpha_1$ and $\alpha_2$, which (due mainly to Lunar ranging 
experiments) are subject to constraints~\cite{will} given by 
\begin{equation}\label{limal}
\alpha_1<2\times 10^{-4}\, , \qquad
\alpha_2<4\times10^{-7}\, .
\end{equation}
In order to be compatible with these limits also,
in addition to satisfying the limit (\ref{123a}), the unperturbed 
extrinsic curvature must satisfy the restriction
\begin{equation}\label{123b}
\frac{K_{_0a}}{K_{_{00}}}\approx -{1\over 4} \alpha_1 w_a
\end{equation}
with $\alpha_1$ subject to (\ref{limal}) where the velocity vector
$w_a$ has magnitude of the order of 200 km\,s$^{-1}$ while
its direction is opposite to that of the cosmic microwave background
relative to the Earth. 

Before considering the further restrictions imposed by global
cosmological considerations on this model for effective gravity in a
brane-world scenario, we shall briefly digress to discuss analogous
local effects in 2-branes.

\subsection{Application to the case $p=3$}

As a second application, consider the well known ``museum'' experiment
used to illustrate how gravity is a consequence of spacetime geometry
and in which a heavy bead of mass $M$ is placed on a piece of fabric
of tension, $\tn$ say, in the local
terrestrial gravitational acceleration field with magnitude $g$. The
weight of the heavy bead produces a dip in the level of the
surrounding fabric, and thus creates an effective potential well in
which another lighter bead may be observed to orbit the central point
of attraction in much the same way as a planet orbits the Sun, but
with an acceleration that varies inversely as the distance, not
inversely as the square of the distance as in the Keplerian planetary
case. In this example, at lowest order the component of the force
density orthogonal to the worldsheet (vertical in this case) is
\begin{equation}
\overline f=\tn g \,,
\end{equation}
and the corresponding displacement will be given by
\begin{equation}
\xi=\frac{g}{2\pi\tn}M\ln r \,,
\end{equation}
so that the effective gravitational potential is
\begin{equation}
V_{_{\rm eff}}=\frac{g^2}{2\pi\tn}M\ln r \,,
\end{equation}
as long as we are sufficiently far from the centre for 
the weak-field regime description to be applicable.  On approaching
the centre, on enters a strong-field regime.

A less trivial but much more speculative application is implicitly suggested 
by the recent proposal~\cite{dgs01} that the rotation curve of spiral
galaxies may be explained by supposing that these galaxies are trapped on
lightweight domain wall, of which the thickness is comparable 
with the thickness of the luminous part of the galaxy. The idea 
of Dvali {\em et al} is that the domain wall supports a massless scalar 
particle zero mode confined to (2+1)--dimensions with a coupling to
the galactic matter so that at distances larger than the typical thickness 
of the wall the total potential for the net force that acts will have an 
extra logarithmic contribution. The formula
\begin{equation}
v^2=r\frac{\dd V}{\dd r}.
\end{equation}
can be used to obtain an observational estimate of the radial dependence
of the effective gravitational potential $V$ from measurements of the
galactic rotation velocity $v$. In typical cases the result can be described
in terms of a plateau at distances larger than the radius of the galaxy
disk where $v$ tends towards a constant limit $v_{_\infty}$ so that
one obtains an empirical formula of the form
\begin{equation}\label{Dvali}
V=-\frac{{\rm G} M_{_{\rm Gal}}}{r}+v_{_\infty}^{\, 2}\ln{r} \,,
\end{equation}
with $v$ ranging from  $10^{-4}$ to $10^{-3}$ in the relativistic units we 
are using, that is $v_{_\infty}$ ranges from 60~km\,s$^{-1}$ to 300~km\,s$^{-1}$.
It was suggested in~\cite{dgs01} that these observations be used to tune the
parameters, such as the energy scale, of their light scalar field,
which does not imply any cosmological catastrophe.

What we wish to point out here is that this scalar field mechanism
is very similar in its effect to the simulated gravity mechanism we
have been describing here.
According to our preceding analysis, if the galaxies were
constrained to move on a 2-brane  subject to an acceleration 
corresponding to a local extrinsic curvature  component
$K_{_{00}}$ then, in addition to the genuinely gravitational
inverse square law contribution, there would be an extra 
contribution to the effective gravitational field so that
the effective gravitational potential would be given by
\begin{equation}
V_{_{\rm eff}}=-\frac{{\rm G}M_{_{\rm Gal}}}{r}+\frac{K_{_{00}}^{\,2}}
{\pi\tn}
M_{_{\rm Gal}}\ln{r},
\end{equation}
This would evidently have the same effect as the potential
(\ref{Dvali}) if the acceleration and tension of the membrane satisfy
\begin{equation}
\frac{K_{_{00}}^{\,2}}{\tn}=\frac{\pi v_{_\infty}^{\,2}}
{M_{_{\rm Gal}}}\, .
\end{equation}
Of course, in order for this variant of the original mechanism to 
be made astrophysically
plausible, it would still be necessary to find some natural mechanism for 
actually trapping the galaxies on the 2-brane. 

\section{Cosmological matching conditions.}
\label{V}
\subsection{Effective Friedmann equation}

It should be remarked that none of the preceding results depend on the
specific functional form of the background metric coefficient ${\cal V}$
in (\ref{50}).  The most general solution, as derived in~\cite{BCG}
and discussed in the preceding work~\cite{CU01,Us01} is
\begin{equation}\label{120} 
{\cal V}=k-{{2\Lambda} \over {p(p-1)}}r^2-{2{\rm G}_{[p+1]}{\cal M}
\over r^{p-2}}\, 
\end{equation}
The first term is the curvature, $k$, of the homogeneous isotropic
space metric $\dd\ell^2$, the second term is the effect of the
cosmological constant, $\Lambda$, of the bulk and the third term is
the contribution of the total asymptotic mass, $\cal M$.
Substituting this in (\ref{52}), and using the solution (\ref{70}) of
the worldsheet evolution equation (\ref{40}), we immediately obtain
the relevant Friedmann type equation for the comoving lengthscale
$r=a$ of the brane in the form 
\begin{equation}\label{122} 
\left({\dot a\over a}\right)^2= {1\over{\cal U}^2}\left(
{\overline f\over p}+ {E_{_{\rm B}}\over a^{p}}\right)^2 
+ {2\Lambda\over{p(p-1)}}+\frac{2{\rm G}_{[p+1]}\cal M}{a^{p}}
-\frac{k}{a^2}\, .
\end{equation}
As observational evidence suggests that the spatial curvature of the
universe does not play a significant role, we shall assume that $k=0$
from now on.  When the gravitational coupling is very small, the
asymptotic mass would have to be very large to have a significant
effect so we shall also set ${\cal M}=0$.  However, in the case where
there is no gravity, this term could be reinstated with a different
physical interpretation.

In order to match our model to what is observationally known about the
large-scale structures of our actual universe, an obvious first step
is to set the brane worldsheet dimension to $p=4$.  As in the
preceding discussion~\cite{CU01} of the gravitationally coupled case,
the low density epoch in which we exist can be dealt with by
an expansion in powers of the dimensionless density ratio
\begin{equation}\label{126}
\varepsilon={\overline{\rho}\over\tn} \, .
\end{equation}
Equation (\ref{122}) can be rewritten as
\begin{equation}\label{300}
\left(\frac{\dot a}{a}\right)^2=\frac{\overline\Lambda_{_4}}{3} +
\frac{1}{\tn^2}\left[
\frac{1}{(1+\varepsilon)^2}\left(\frac{\overline f}{4}+
\frac{E_{_{\rm B}}}{a^4}\right)^2-\left(\frac{\overline f}{4}\right)^2
\right]
\end{equation}
where the effective 4-dimensional cosmological constant,
$\overline\Lambda_{_4}$, is given in terms of its 5-dimensional analogue,
$\Lambda$, and the force, $\overline f$, by
\begin{equation}\label{132} 
{\overline\Lambda_{_4}\over 3}={\Lambda\over 6}+\left({\overline f\over
4\tn}\right)^2\, .
\end{equation}
It can immediately be verified that (\ref{132}) agrees with what is
obtained from the corresponding formula in the preceding
analysis~\cite{CU01} when the background gravitational coupling constant
is set to zero. However, as we have already pointed out, if we similarly set the background gravitational
coupling to zero in the corresponding formula~\cite{CU01} for the
effective 4-dimensional gravitational coupling constant ${\rm G}_{_{[4]}}$ we
would obtain a negative value that cannot possibly matched to the Newton
constant ${\rm G}_{_{\rm N}}$ that is actually observed.  This is because the
scheme of the previous work corresponds to assuming that the constant
$E_{_{\rm B}}$ is sufficiently small for its contribution to be entirely
neglected in the late time limit considered here, so that the bracketed
contribution in (\ref{300}) contains at lowest order just the
$\varepsilon$ term, which has the desired property of proportionality to
$\rho$, but with the wrong sign.

The idea of the present approach is to get round this apparent 
obstacle to taking the zero background gravity limit by making an
essentially different hypothesis, namely that instead of being
small, $E_{_{\rm B}}$ is large enough to give a positive contribution
that can overcome the negative contribution proportional to 
$-\varepsilon$. The cosmological scenario will thus have an era 
dominated by what we refer to as pseudo-radiation, that will last until 
at least the present epoch  the unless it was superseded by a 
transition (which if it occurred must have been recent) to an epoch of 
domination by a cosmological constant. Instead of the term ``mirage matter'' 
that was used by Kehagias and Kiritsis~\cite{move1} whose pioneering 
analysis of gravitationally uncoupled brane-world scenarios originally 
drew attention to the potential cosmological importance of such an 
effect, we prefer to use the term ``pseudo-radiation'' for the background 
geometric contribution dominating the effective Friedmann equation because, 
like a true radiation gas, its energy density evolves proportionally to 
$a^{-4}$, while unlike an ordinary mirage its dynamical effect is very real 
-- in our case actually preponderant.

\subsection{The nucleosynthesis condition}

 In order for the kind of scenario described in the preceding subsection
to be realistic,  we need {\it at least} to require that
the four parameters ($\Lambda,\tn,\overline f,E_{_{\rm B}}$)
are such that at nucleosynthesis, the pseudo-radiation term dominates the
Friedmann equation (in particular compared with terms in $a^{-8}$) and
that its amplitude has the value required to get the correct
nuclear abundances, i.e. when the temperature, $\Theta$, is comparable
with the nuclear reaction value, $\Theta_{_{\rm N}}\approx 1$ MeV,
the expansion rate must be given approximately by
\begin{equation}\label{Newt}
\left(\frac{\dot a}{a}\right)^2\simeq {8\pi {\rm G}_{_{\rm N}}
\over 3}\rho_{_{\rm r}}\, ,\end{equation}
where $\rho_{_{\rm r}}$ is the radiation density as given by the
formula
\begin{equation}\label{rom}
\rho_{_{\rm r}}=\frac{\pi^2}{30}g_*\Theta^4\,,
\end{equation}
in which the value of the number $g_*$ of effective relativistic degrees
of freedom is calculated to allow for the photon field and three
families of degenerate neutrinos which gives $g_*=10.75$.
In order to avoid getting a negative value for the right hand side
of (\ref{300}) at some subsequent time,
the parameters must be such that the pseudo radiation contribution
proportional to $a^{-4}$, or equivalently to $\Theta^4$, must be larger
than the true radiation contribution given by (\ref{rom}) and also
than the matter contribution $\rho_{_{\rm m}}$ in the decomposition
$\rho=\rho_{_{\rm m}}+\rho_{_{\rm r}}$
In order to be consistent with the standard picture of development of clustering from small
perturbations  the cosmological constant must remain subdominant
until at least the present epoch.

To determine the implication from these requirements, we expand (\ref{300}) 
to get
\begin{equation}\label{301}
\left(\frac{\dot a}{a}\right)^2=\frac{\overline\Lambda_{_4}}{3} +
\frac{1}{\tn^2}\left[\frac{\overline f}{2}
\frac{E_{_{\rm B}}}{a^4}+\frac{E^2_{_{\rm B}}}{a^8}-2\varepsilon
\left(\frac{\overline f}{4}+
\frac{E_{_{\rm B}}}{a^4}\right)^2
\right]+{\cal O}(\varepsilon^2)
\end{equation}
and make the decomposition
$\varepsilon=\varepsilon_{_{\rm m}}+\varepsilon_{_{\rm r}}$,
where $\varepsilon_{_{\rm m}}={\rho_{_{\rm m}}/\tn}$
and $\varepsilon_{_{\rm r}}={\rho_{_{\rm r}}/\tn}$.
These terms will be given in terms of their present values by
$\varepsilon_{_{\rm m}}=\varepsilon_{_{\rm m}}^0\vartheta^3$ and
$\varepsilon_{_{\rm r}}=\varepsilon_{_{\rm r}}^0\vartheta^4$,
in terms of the reduced temperature, as defined in terms of
its present value $\Theta_0$ or the present value $a_0$ of the cosmological
scale factor by
\begin{equation}\label{304}
\vartheta=\frac{\Theta}{\Theta_0}=\frac{a_0}{a}\, .
\end{equation}
The equation (\ref{301}) is thereby expressible in the form,
after setting $a_0=1$,
\begin{eqnarray}\label{305}
\left(\frac{\dot a}{a}\right)^2&=&\frac{\overline\Lambda_{_4}}{3} +
\frac{1}{\tn^2}\left[ \frac{\overline f}{2}\left(E_{_{\rm
B}}-\frac{\overline f}{4}\varepsilon_{_{\rm
r}}^0\left(1+\frac{\vartheta_{_{\rm
eq}}}{\vartheta}\right)\right)\vartheta^4 +E_{_{\rm B}}\left(E_{_{\rm
B}}-{\overline f}\varepsilon_{_{\rm r}}^0\left(1+\frac{\vartheta_{_{\rm
eq}}}{\vartheta}\right)\right)\vartheta^8 \right.\nonumber\\
&&\qquad\qquad\qquad\left.-2E_{_{\rm
B}}^2\varepsilon_{_{\rm r}}^0\left(1+\frac{\vartheta_{_{\rm
eq}}}{\vartheta}\right)\vartheta^{12}\right]+{\cal O}(\varepsilon^2)
\end{eqnarray}
where we have introduced the matter-radiation density equality 
value
\begin{equation}
\vartheta_{_{\rm eq}}\equiv\frac{\varepsilon_{_{\rm m}}^0}{
\varepsilon_{_{\rm r}}^0}\, ,
\end{equation}
of the reduced temperature, which is much larger than unity 
($\vartheta_{_{\rm eq}}\sim10^3$ in standard cosmology) and in terms
of which the total matter component takes the form
\begin{equation}
\varepsilon=\varepsilon_{_{\rm r}}^0\vartheta^3
(\vartheta+\vartheta_{_{\rm eq}})\, .
\end{equation}
Note that the terms of ${\cal O}(\varepsilon^2)$ in (\ref{305}),
although containing terms of order $\vartheta^6$, are suppressed by an
additional factor of $\varepsilon_{_{\rm r}}^0$ and thus can be
dropped legitimately (see below).

In order for the term proportional to $\vartheta^4$ to remain positive until 
at least the present time, $\vartheta\simeq 1$, we evidently
need that the pressureless matter term be smaller that the pseudo
radiation term proportional to $E_{_{\rm B}}$, giving the constraint
\begin{equation}\label{306}
4E_{_{\rm B}}>\overline f\varepsilon_{_{\rm r}}^0\vartheta_{_{\rm eq}}
\end{equation}
and furthermore, to avoid premature domination by the cosmological constant,
we also need
\begin{equation}\label{307}
\overline f\big(4E_{_{\rm B}}-\overline f\varepsilon_{_{\rm r}}^0
\vartheta_{_{\rm eq}}\big)\geq
\frac{8}{3}|\Lambda_{_4}| \tn^2 \,.
\end{equation}
This implies that the term in $\vartheta^3$, which has the wrong sign, must
{\it always} have been subdominant until now: it can {\it at most} 
equal the pseudo-radiation component today, and thus was
much less in the past since it redshifts more slowly.

Since the true radiation term is always a fixed fraction of the 
pseudo-radiation term, the fact that our scenario is compatible with the
nucleosynthesis data leads first to the identification of the Newton
constant, ${\rm G}_{_{\rm N}}$ in the formula (\ref{Newt}) as  
\begin{equation}\label{308}
{\rm G}_{_{\rm N}}=\frac{3\overline f}{16\pi \tn^3}
\left(\frac{E_{_{\rm B}}}{\varepsilon_{_{\rm r}}^0}
-\frac{\overline f}{4}\right).
\end{equation} 
Since $\vartheta_{_{\rm eq}}\gg 1$,
the positivity of ${\rm G}_{_{\rm N}}$ is evidently ensured by (\ref{306}).
It is to be observed that this globally determined quantity 
${\rm G}_{_{\rm N}}$ is {\it a priori} different from the quantity 
${\rm G}_{_{[4]}}$ determined according to (\ref{118}) by local physics, 
so the requirement that they should ultimately agree will entail
a rather excessive amount of fine tuning.

In order for the term in (\ref{305}) proportional to $\vartheta^8$ to be 
negligible compared to the one proportional to $\vartheta^4$ at the epoch of
nucleosynthesis we evidently  need to have
\begin{equation}\label{309}
\overline f\big(4E_{_{\rm B}}-\overline f\varepsilon_{_{\rm r}}^0\big) >
8E_{_{\rm B}}\big(E_{_{\rm B}}-\overline f\varepsilon_{_{\rm r}}^0
\big)\vartheta_{_{\rm N}}^4
\end{equation}
where $\vartheta_{_{\rm N}}\sim 10^{10}$ is the value of $\vartheta$
at nucleosynthesis. In conjunction with (\ref{306}), this latter 
constraint simply gives the requirement that the four parameters
$E_{_{\rm B}}$, $\overline f$, 
$\tn$ and $\Lambda$ should satisfy the conditions
expressible in terms of the present cosmological radiation density
\begin{equation} \rho_{_{\rm r}}^0=\tn\varepsilon_{_{\rm r}}^0
={\pi^2\over 30} g_\ast \Theta_0^{\, 4}\, ,\end{equation}
as the inequalities
\begin{equation}\label{311}
\vartheta_{_{\rm N}}^{-4}\gg 2\frac{E_{_{\rm B}}}{\overline f}>
{\rho_{_{\rm r}}^0\vartheta_{_{\rm eq}}\over \tn}
\, .\end{equation} 
The only other restrictions on the four listed parameters are the
conditions obtained from the inequality (\ref{307}) and the
identification (\ref{308}) --- in which the second term on the right
is negligible by (\ref{306}) --- so that we are left with 
\begin{equation}\label{Newto} 
{16\pi\rho_{_{\rm r}}^0\over 3m_{_{\rm P}}^2}
\simeq
{\overline fE_{_{\rm B}}\over \tn^2}\gta
{2\over 3}\overline\Lambda_{_4}\, ,
\end{equation}
using the standard definition, $m_{_{\rm P}}\equiv {\rm G}_{_{\rm N}}^{-1/2}$,
of the Planck mass, whose numerical value is given by
$m_{_{\rm P}}\approx 10^{22}$ MeV. 
No further restriction is needed to ensure 
that the term proportional to $\vartheta^{12}$ in (\ref{301})
remains negligible at temperatures up to the nucleosynthesis range.

It can now be seen to follow directly from (\ref{311}) that
the brane mass scale $m_{_\infty}$, defined by $m_{_\infty}^4\equiv\tn$,
must satisfy
\begin{equation}\label{312}
m_{_\infty}\gg\left(\rho_{_{\rm r}}^0\vartheta_{_{\rm eq}}\right)^{1/4}
\vartheta_{_{\rm N}}.
\end{equation}
With the standard values $\vartheta_{_{\rm eq}}\approx 10^3$,
$\vartheta_{_{\rm N}}\approx10^{10}$
and  $\Theta_0\approx
 2\times10^{-10}$~MeV, which corresponds to
$\rho_{_{\rm r}}^0\sim6\times10^{-39}\,(\hbox{MeV})^4$, we get that
\begin{equation} \label{minf}
m_{_\infty}\gg 10\,\hbox{MeV}
\end{equation}
which is comparable with what we found in our previous
work~\cite{CU01}. The bound (\ref{minf}) implies that
$\varepsilon_{_{\rm r}}^0\ll10^{-44}$ which justify {\em a posteriori}
that we can neglect terms of order $\varepsilon^2$ in (\ref{305}).

The first equality in (\ref{Newto}) will now give an order of magnitude 
inequality for the product $\overline fE_{_{\rm B}}$, which will
be given in terms of the brane tension by
\begin{equation} \label{Lal}
{\overline fE_{_{\rm B}}\over\tn^2}\approx  10^{-81} (\hbox{MeV})^2\, ,
\end{equation}
and hence by (\ref{minf}) will satisfy $\overline fE_{_{\rm B}}\gta
10^{-73} (\hbox{MeV})^{10}$. In order for the cosmological constant to
have become important by the present epoch (as suggested by some
recent supernova data) we would need that the
ratio $\overline\Lambda_{_4}/H_0^{\,2}$ should be comparable with unity 
where $H_0$ is the present day value of the Hubble expansion rate
which is given roughly by where $H_0\approx 10^{-38}\,\hbox{MeV}$.
However it follows from (\ref{Newto}) that this condition would fail
by a considerable margin 
since substitution of (\ref{Lal}) gives 
\begin{equation}
\overline\Lambda_{_4}/H_0^{\, 2} \lta 10^{-11}
\, .\end{equation}
The limits on $E_{_{\rm B}}$ obtained from (\ref{311}) by substitution
from the first  equality in (\ref{Newto}) have the form

\begin{equation}
10^{-56}
\left(\frac{m_{_\infty}}{10\,\hbox{MeV}}\right)^4\,\hbox{MeV}^{5}\gg
E_{_{\rm B}}>10^{-56}
\left(\frac{m_{_\infty}}{10\,\hbox{MeV}}\right)^2\,\hbox{MeV}^{5}.
\end{equation}

\section{Discussion}\label{VI}

It seems from the former paragraph that we can find a set of
parameters for our model to be compatible with large-scale cosmology
(that is, with nucleosynthesis and a cosmological constant dominated
universe today), even if very contrived. But it has to be emphasized
that the large-scale  value of the gravitational constant
\begin{equation}\label{globG}
{\rm G}_{_{\rm N}}=\frac{3}{8\pi\tn^3}\frac{
\overline f}{2}\left(\frac{E_{_{\rm B}}}{\varepsilon_{_{\rm r}}^0}
-\frac{\overline f}{4}\right)
\end{equation}
has no reason {\it a priori} to agree with its small 
scale value (\ref{118})
\begin{equation}\label{locG}
{\rm G}_{_{[4]}}=\frac{(K_{_{00}}^{_{(\rm local)}})^2}{4\pi\tn}.
\end{equation}

We first investigate here, the compatibility between the local and
global properties of the gravitation and then discuss the viability of
such a model as well as its possible extensions.

\subsection{Compatibility between local and global matching}

To compare the two values of the Newton constant, respectively
determined from the local Newton law and global cosmology, we need to
compute the extrinsic curvature in the cosmological framework. From
(\ref{72}) we get,  at first order in $\varepsilon$, that the extrinsic
curvature scalar is given by
\begin{equation}\label{400}
K^{_{\rm(global)}}=-\frac{\overline f}{\tn}+\left[
\frac{6+\Gamma}{4}\overline f+3\Gamma\,E_{_{\rm B}}\vartheta^4
\right]\frac{\varepsilon}{\tn}+
{\cal O}(\varepsilon^2)\, ,
\end{equation}
on the assumption that, as described in more detail in the appendix,
one assumes a polytropic equation of state with index $\Gamma$ for which
the pressure of the cosmic fluid in (\ref{76}) will be given by 
$P=(\Gamma-1)\rho$. Equation (\ref{56}) then implies that
we shall have 
\begin{equation}
K_{_{00}}^{_{\rm(global)}}=\frac{\overline f}{4\tn}-
\frac{3}{\tn} E_{_{\rm B}}\vartheta^4
-\left[\overline f-3(\Gamma+1)
\left(\frac{\overline f}{4}+E_{_{\rm B}}\vartheta^4
\right)\right]\frac{\varepsilon}{\tn}+ {\cal
O}(\varepsilon^2).
\end{equation}
In the relevant range (\ref{311}) for the admissible parammeter values,
the large-scale value of the extrinsic curvature components will
therefore be given at lowest order in $\varepsilon$ today (i.e. when 
$\vartheta\sim1$) by
\begin{equation}\label{410}
K^{_{\rm(global)}}\simeq-\frac{\overline f}{\tn}\,,\qquad
K^{_{\rm(global)}}_{_{00}}\simeq-\frac{1}{\tn}\frac{\overline f}{4}\, .
\end{equation}
The extrinsic curvature vector will thus be subject to what is
interpretable as a space-time isotropy (Lorentz invariance)
condition, namely  
\begin{equation}\label{alpha} K^{_{\rm(global)}}_{_{00}}
= -\frac{1}{4}K^{_{\rm(global)}}\, ,
\end{equation}
which is evidently quite different from  the
observational constraint (\ref{123a}). However the apparent contradiction
between these two conditions need not be of any consequence if,
as we have already concluded, the relevant local and global
curvature values are independent.

To reconcile the cosmologically determined
value ${\rm G}_{_{\rm N}}$ given by (\ref{globG}) with the locally
measured value ${\rm G}_{_{[4]}}$, given by (\ref{locG}), of the 
gravitational constant we need to have 
\begin{equation}
K_{_{00}}^{_{\rm(local)}}\simeq\pm\left(\frac{3\overline f}{4\tn^2}
\frac{E_{_{\rm B}}}{\varepsilon^0_{_{\rm r}}}\right)^{1/2} \,,
\end{equation}
with the supplementary requirement 
$K^{_{\rm(local)}}=2K_{_{00}}^{_{\rm(local)}}$.
This local value is obviously different from the value (\ref{410})
found on large scales since we shall have 
\begin{equation}
\left|K^{_{\rm(global)}}_{_{00}}\right|=\alpha
\left|K_{00}^{_{\rm(local)}}\right|\,\qquad{\rm with}\qquad
\alpha^2\equiv12\varepsilon^0_{_{\rm r}}\frac{\overline f}{E_{_{\rm B}}}
\, ,
\end{equation}
where by (\ref{311}), we get that the parameter $\alpha$ is in the range
\begin{equation}
2\times10^{-3}\gsim\alpha^2\gsim9\times10^{-3}\left(
\frac{m_{_\infty}}{10\,{\rm Mev}}\right)^{-4}
\end{equation}
from which it follows, a fortiorei,  that $\alpha<1$.

Such mismatch between the small-scale and large-scale values of the
extrinsic curvature does not seem unreasonable in view  the consideration 
that the background configuration of brane  does not need to be strictly
homogeneous and isotropic, as assumed until now, but may have different
curvature on small and large scales. The cosmological constraints
determine the large-scale extrinsic curvature, while light
deflection observations so far give precise information only about local
extrinsic curvature. This raises the question of the transition scale
between these two regimes which should range somehow above the galactic
scale. This also leads to a potentially lethal test for our model: since
the local extrinsic curvature has no reason to be the same in two
different galaxies, the measurement of the light deflection should lead
to different results in different galaxies.

\subsection{Viability of the model}

We have shown that as well as automatically providing a local
gravitational interaction of the familiar Newtonian kind, 
the simulated gravity model presented here can be contrived in such
a way as to satisfy the most stringent cosmological constraint 
 -- namely the one provided by nucleosynthesis. However it is still
confronted with other serious problems.

\begin{itemize}
\item
Firstly, it suffers from a double fine tuning requirement
arising from the fact that, on
local scales, the gravitational constant depends a priori on time. We
must require not only that the weak-field limit  be instantaneously
compatible with the
the condition (\ref{123a}) for consistency with the
observed deflection of light, but also that the predicted
gravitational constant (\ref{118}) should vary slowly enough to be
consistent with the experimental limit~\cite{gdot}
\begin{equation}\label{gdotmax}
\left|\frac{\dot {\rm G}_{_{[4]}}}{{\rm G}_{_{[4]}}}
\right|\leq 6\times10^{-12}\,\hbox{yr}^{-1}.
\end{equation}
\item Secondly, the predicted post-Newtonian effects and the
current observational limits~\cite{will} imply that the local value
extrinsic curvature is subject not just to the limitation
its mixed components $K_{_0a}$ given by (\ref{limal}) and
(\ref{123b}) but on the further fine tuning condition
subjecting its time component $K_{_{00}}$ to (\ref{123a}), which
is eqivalent to the condition that the extrinsic curvature 
should turn out to be almost exactly of the form 
\begin{equation}\label{com1}
 K_{\mu\nu}=K_{_{00}}\big(2\overline
u_\mu\overline u_\nu +\overline\gamma_{\mu\nu}\big)\,,
\end{equation} 
entailing  $\nu'\sim -r'/a$ in (\ref{56}). Note that, at this stage
of our investigation, these restrictions are not worse than those
occuring e.g. for scalar-tensor theories of gravity (see~\cite{br} for
a comparison) before the mechanism of attraction~\cite{dn} toward General
Relativity was discovered. However at this stage we are not aware of any
natural mechanism that would lead to the form (\ref{com1}) satisfying 
the post-Newtonian constraints as a preferred outcome.

In addition to the kinds of observational restriction we have
considered so far, it is to be noted that there will be others involving
non-linear effects and motion of the source. It is not our purpose to
investigate all these constraints here, but just to point out that they
are of potential importance whenever alternative theories of gravity
are under consideration, and that merely recovering the standard Newton
law is in general not enough. None of these post-Newtonian constraints
have been explicitly taken into account so far in this or other works
on brane-world scenarios. However, in typical brane-world models, (even
those of asymmetric type) gravity at Solar system scales is describable
in terms of a spin 2 graviton field, and in such cases it can be
expected that the post-Newtonian constraint will automatically be
satisfied.

\item Thirdly, concerning the cosmology of our model, the expansion
rate switches from a pseudo-radiation to a cosmological constant
dominated phase whereas the matter content switches from a radiation
dominated era to a matter dominated era (at $\vartheta_{_{\rm eq}}$)
and then to a cosmological constant dominated era (at
$\vartheta\sim1$). This is a non-standard cosmological scenario that 
does not have an ordinary matter-dominated era characterised by a 
Friedmann type evolution equation of the form $H^2\propto a^3$.  
raises the question of how  large-scale structure will develop in 
this kind of scenario. In the case of the standard cosmology, it is 
known that density perturbations needed for subsequent galaxy formation 
cannot grow during a radiation dominated era. However it is not
yet clear what may happen in a pseudo radiation era in which the
strength of gravity on shorter scales is affected by the nature of
fluctuations on longer scales.
\item
Fourthly, in relation to the preceeding remark, we need a transition 
scale between the local and global behaviour that has to range a 
somewhere above the galaxy scale. However, this scale is not \
directly determined  by the model and might involve a third fine tuning.
A potentially relevant test that can be thought of is weak gravitational 
lensing on cosmological scales~\cite{ub00} which can provide a probe of 
the Newtonian law of gravitation (and more particularly of the Poisson
equation) up to some hundreds of Megaparsec when data are
available. This may be able to provide  constraints on this transition
scale.
\end{itemize}

\subsection{Conclusions}

In this article, we have considered a simulated gravity model when the
reflection symmetry is broken in brane-world scenarios for which
genuine (bulk) gravitational coupling is absent, but in which the
possible effect of a gauge four form coupling is allowed for.  A
previous investigation by Kehagias and Kiritsis~\cite{move1} of this
gravitationally decoupled ``probe'' limit\footnote{
The motion of these they sometimes describe as ``geodesic motion of a
probe D3-brane'', which is potentially misleading as the evolution is
not generated by geodesics in the usual sense of the term.}
is noteworthy for having drawn attention to the dynamical
importance of what they referred to as ``mirage'' matter, meaning the
contribution we have preferred to describe as pseudo-radiation. 
Their work was restricted to homogeneous isotropic cosmological
configurations of any codimension, and allowed only for the kinds of
matter field most directly motivated by higher dimensional superstring
theory, namely a Born-Infeld type contribution in the brane and a
Ramond-Ramond gauge field outside. 
In this investigation we restrict attention to the codimension one
case but progress further in two ways. The first advance is by allowing
for more realistic matter fields on the brane, and showing that the
so-called ``mirage'' contribution, which we have called
pseudo-radiation, can be adjusted to give consistency with
nucleosynthesis --- the most empirically sensitive cosmological
constraint. The second kind of progress made here is the less
mathematically trivial step of allowing for local deviations from
homogeneity and the demonstration that in the weak-field, small-scale
limit there will be an effective ``simulated'' gravity mechanism with
a Newtonian type inverse square law behaviour.

The outcome of the first advance is rather satisfactory: having
derived the equation of evolution of the universe we show that, for a
certain range of parameters, it can be made compatible with
nucleosynthesis subject to the proviso that the cosmological constant
is not yet dominant today. However the outcome of the second kind of
progress raises as many problems as it solves: as discussed in the
previous section, agreement at the local level can only be obtained by
adjusting the relevant parameters in an artificially contrived
manner. It necessitates at least two fine tunings (arising from the
time variation of the gravitational constant and of the compatibility
with the post-Newtonian limit), and requires an undetermined
transition lengthscale between local and global properties of our
3--brane. Let us emphasize that it is not enough to recover Newton
gravity on local scales and to produce a viable cosmology; one also
needs to take the post-Newtonian constraints into account.  Further
work would be needed to derive the detailed evolution of density
fluctuations in our model, but it does not seem that it would lead to
an acceptable description of structure formation, at least if the
standard picture is still valid (which is not entirely obvious).

Different, more sophisticated, scenarios might be conceived, in which
one might try to overcome the problems that have been pointed out by
allowing more general fields to act in the bulk.  Another possibility
is to change the background geometry, which we can do without changing
the bulk matter content when there is no gravity.  By tuning the
background, it is very likely that we could get a cosmology
with the standard Friedmann law.  In any case, even if the effective
gravity mechanism described here is not dominant, it will always be
present to some extent as a contributing effect in reflection symmetry
breaking brane-world scenarios. Whereas most previous work has
concentrated on strongly self gravitating reflection symmetric case,
the present article initiates investigation of the opposite extreme
limiting case of brane-worlds that are very weakly self gravitating
but for which gravity is simulated by the effect of reflection
symmetry violation.

\section*{Acknowledgements}

We wish to thank Daniele Steer for interesting discussions on moving branes,
and Gilles Esposito-Far\`ese for discussions on scalar-tensor theories.
We have also benefitted from informative discussions with Jos\'e Blanco-Pillado
and Martin Bucher about previous relevant work~\cite{gv} on 2-branes. We are 
grateful for support from  P.P.A.R.C. (for R.A.B.) and from E.P.S.R.C. 
(for A.M.).

\appendix
\section{Variational formulation in static background}

The variational principle for a minimally coupled 3-brane in a
given background specified by a metric $g_{\mu\nu}$ and a 
(Ramond-Ramond type~\cite{move1}) gauge 4 form 
$A_{\mu\nu\rho\sigma}$ will be given~\cite{BC01} 
in terms of a Lagrangian density ${\cal L}$ depending only
on internal fields on the brane worldsheet, and of a ``brawn'' coupling
(generalised charge) constant $e_{_{[4]}}$,  by an action integral
of the form
\begin{equation}\label{A1}
{\cal I}=\int \big({\cal L}+{e_{_{[4]}}\over 4!}{\cal E}^{\mu\nu\rho\sigma}
A_{\mu\nu\rho\sigma}\big)\, \dd\overline{\cal S}\, ,\end{equation}
where $\dd\overline{\cal S}$ is the element of 
worldsheet volume measure as given in terms of internal coordinates 
$\sigma^i$,
($i$=0,1,2,3) and the determinant $|\ivm|$ of the induced
metric $\ivm_{ij}$ by 
\begin{equation} \dd\overline{\cal S}=\Vert\ivm\Vert^{1/2}\,
\dd\sigma^{_0}\,\dd\sigma^{_1}\,\dd\sigma^{_2}\,\dd\sigma^{_3}
\, ,\end{equation}
and where ${\cal E}^{\mu\nu\rho\sigma}$ is the corresponding
antisymmetric unit worldsheet tangent tensor (normalised so that
${\cal E}^{\mu\nu\rho\sigma}{\cal E}_{\mu\nu\rho\sigma}=4!$)
as specified in terms of the associated internal measure
tensor ${\cal E}_{hijk}$ by 
\begin{equation} {\cal E}^{\mu\nu\rho\sigma}={\cal E}^{hijk}\,
x^\mu_{\, ,h}\, x^\nu_{\, ,i}\, x^\rho_{\, ,j}\, x^\sigma_{\, ,k}
\, .\end{equation}

For a bulk metric of the static form (\ref{50}) the
space metric $\dd\hat s^2=r^3\,\dd \ell^2$ of the homogeneous 
isotropic space sections will similarly determine an analogous 
antisymmetric tensor $\hat {\cal E}^{\mu\nu\rho}$ that is tangent to 
the homogeneous isotropic space 3-sections with normalisation 
$\hat {\cal E}^{\mu\nu\rho}\hat {\cal E}_{\mu\nu\rho} =6$.
This space (as opposed to space time) measure tensor can be used
to define a conformally related tensor
\begin{equation} {\cal Y}_{\mu\nu\rho}
= r\,\hat{\cal E}_{\mu\nu\rho}\, ,\end{equation}
with the noteworthy property of having a covariant derivative that
is entirely antisymmetric, which qualifies it for description 
as a Killing Yano tensor. This special property can be seen to follow
just from the ``warped product'' form (\ref{50})
of the full spacetime metric $\dd s^2$ (independently of the special 
homogeneity and isotropy property postulated for $\dd\ell^2$)
which enables us to evaluate the covariant derivative as
\begin{equation} \nabla_{\!\mu} {\cal Y}_{\nu\rho\sigma}
={4\over r} r_{[,\mu} {\cal Y}_{\nu\rho\sigma]}\, .\end{equation}
It follows that the symmetric product tensor
\begin{equation} {\cal K}_{\mu\nu}={1\over 2}{\cal Y}_{\mu\rho\sigma}
{\cal Y}_\nu^{\ \rho\sigma}\, ,\end{equation}
with trace ${\cal K}_\mu^{\ \mu}= 3\, r^2$, will be an ordinary Killing tensor,
meaning that its symmetrised derivative vanishes,
\begin{equation}
\nabla_{(\mu}{\cal K}_{\nu\rho)}=0\, ,\end{equation}
so that for a free particle motion it determines a quadratic constant 
of motion of the kind recently discussed in the general context of 
``warped product'' metrics by Droz-Vincent~\cite{D01}. 

The utility of the Killing Yano tensor ${\cal Y}_{\mu\nu\rho}$ for our 
present purpose is that it can be used to construct an explicit static 
solution for the ``brawn'' gauge field $A_{\mu\nu\rho\sigma}$, which 
may be taken to be given in terms of the gradient of the time 
coordinate $t$ in the expression 
(\ref{50}) for the static bulk metric by
\begin{equation}\label{Asol} A_{\mu\nu\rho\sigma}=F\, t_{[,\mu} 
{\cal Y}_{\nu\rho\sigma]} \, ,\end{equation}
where $F$ is a constant, which will determine the value
of the force density $\overline f$
on the right hand side of the equation of motion (\ref{40}) 
according to the prescription~\cite{BC01}
\begin{equation} \overline f= e{_{[4]}}\, F\, ,\end{equation}
and in terms of which the corresponding
gauge independent field tensor,
\begin{equation}    F_{\mu\nu\rho\sigma\tau}=5\nabla_{[\mu}
A_{\nu\rho\sigma\tau]}\, ,\end{equation}
can be seen to be given simply by
\begin{equation} F_{\mu\nu\rho\sigma\tau}= F\epsilon_{\mu\nu\rho\sigma\tau}
\, ,\end{equation}
where $\epsilon_{\mu\nu\rho\sigma\tau}$ is the background spacetime
measure tensor, which will be expressible in terms of the 
coordinates introduced in (\ref{50}) by
\begin{equation}\epsilon_{\mu\nu\rho\sigma\tau} = 20 \,r_{,[\mu} t_{,\nu}
\hat{\cal E}_{\rho\sigma\tau]}\, .\end{equation}

Let us now specialise, as in \S~\ref{III}, to the case in which
the brane is itself homogeneous and isotropic, so that the
evolution of its worldsheet will be describable simply by specifying
$r$ as a function of $t$. In this case the worldsheet tangent
tensor ${\cal E}^{\mu\nu\rho\sigma}$ will simply be specifiable
in terms of the preferred cosmological unit vector $\overline u^\mu$
introduced in (\ref{56}) by
\begin{equation} {\cal E}^{\mu\nu\rho\sigma} =4\overline u^{[\mu}
\hat{\cal E}^{\nu\rho\sigma]}\, ,\end{equation}
and hence by (\ref{Asol})  the coupling term in (\ref{A1}) will be
given just by
\begin{equation} {e_{_{[4]}}\over 4!}{\cal E}^{\mu\nu\rho\sigma}
A_{\mu\nu\rho\sigma} =    {e_{_{[4]}} F \over 4}\bar u^\mu t_{,\mu}
\, .\end{equation}

In the homogeneous isotropic case (and even more generally
for the treatment of adiabatic cosmological perturbations) 
little generality will be lost by supposing that the matter
on the brane is describable as an irrotational perfect
fluid characterised by an equation of state specifying
the energy density $\bar\rho$ as a function of the number density,
$\overline n$ say of conserved particles, which 
(in the subnuclear energy regime we are concerned with)
can taken to be baryons.  In such a case the 4-dimensional
fluid dynamics will have a particularly simple Lagrangian formulation
in which the only independent variable is a scalar potential, 
$\varphi$ say, with a timelike gradient $\varphi_{,\rho}$
whose magnitude $\bar\mu$, as given by
\begin{equation} \overline\mu^2=
- \bar\gamma^{ij}\varphi_{,i}\varphi_{,j}
\, ,\end{equation}
is to be identified with the relevant chemical potential (i.e.
effective mass per baryon) as defined by
\begin{equation} \bar\mu={\dd\bar\rho\over \dd 
\overline n}\, ,\end{equation}
while the Lagrangian itself can be taken to be the corresponding
pressure function $\overline P$, as obtained in terms of $\overline\mu$ 
by application of the standard prescription
\begin{equation} \bar P=\bar n\bar\mu-
\bar \rho\, .\end{equation}
To adapt this simple perfect fluid model to the context
of a brane-world in a 5-dimensional background, all one has to
do is to identify the Lagrangian density
${\cal L}$ in (\ref{A1}) with the negative of the total brane
tension ${\cal T}$ as given by (\ref{76}), i.e. we take
\begin{equation} {\cal L} = -{\cal T}\, ,\end{equation}
with
\begin{equation} {\cal T} =\tn -\overline P
\, .\end{equation}

In order to be explicit, for typical cosmological purposes
a good description will be provided by an equation of state
of polytropic form given in terms of a fixed mass (per baryon)
$m$, a fixed polytropic index $\Gamma$, and a fixed proportionality
constant $\kappa$ say by
\begin{equation}\bar\rho= \bar\rho_{_{\rm m}}+\bar\rho_{_{\rm r}}\, ,
\end{equation}
with
\begin{equation} \bar\rho_{_{\rm m}}= m\overline n\, ,\qquad
\rho_{_{\rm r}}=\kappa\bar n^{\Gamma}
\, ,\end{equation}
which gives 
\begin{equation} \bar\mu=m+\kappa\Gamma\bar n^{\Gamma-1}\, ,
\qquad \bar P= (\Gamma-1)\big(\bar\rho -m\bar n)
\, ,\end{equation}
so that the required pressure function is finally obtained in the form
\begin{equation}
\bar P=\kappa(\Gamma-1)\Big({\bar\mu-m\over\kappa\Gamma}\Big)
^{\Gamma/(\Gamma-1)}\, .\end{equation}
The simplest relevant possibility is that of a pure radiation gas, as 
characterised by $m=0$ and $\Gamma=4/3$ for which the pressure function 
contribution to the Lagrangian density will just be proportional to
$\big(\bar\gamma^{ij}\varphi_{,i}\varphi_{,j}\big)^2$.

Whatever the form of the equation of state, the four dimensional
surface integral in (\ref{A1}) will be proportional in the homogeneous
isotropic case to a one dimensional integral of the form
\begin{equation} I=\int r^3\Big(-{\cal T}\, \dd\tau+
e_{_{[4]}}\, F{r\over 4}\,\dd t\Big)
 \, .\end{equation}
Using the explicit form of the metric (\ref{50}) to evaluate
the derivative of the proper time $\tau$ with respect to the background 
metric time $t$, this will be expressible in the standard  form
\begin{equation} I=\int L\, \dd t\, ,\end{equation}
for an ordinary Lagrangian (not Lagrangian density) function given by
\begin{equation} L = -{\cal T}\left({\cal V} -{\cal V}^{-1}
\Big({\dd r\over \dd t}\Big)^2\right)^{1/2} r^3 +
e_{_{[4]}}\, F\, {r^4\over 4} \, ,\end{equation}
in which the equation of state specifies $\overline P$,  and hence
also ${\cal T}$,  as a function of $\bar\mu$
and hence of $\dd\varphi/\dd t$ and $\dd r/\dd t$ via the
relation
\begin{equation} \bar\mu={\dd\varphi\over\dd \tau}=
{\dd\varphi\over \dd t}\left({\cal V} -{\cal V}^{-1}
\Big({\dd r\over \dd t}\Big)^2\right)^{-1/2}{\cal V}^{-1/2}
\, .\end{equation}
In terms of the momenta defined by the variation
\begin{equation}\label{momd} \dd L ={\partial L\over\partial r} \dd r
+p_r\dd\Big({\dd r\over \dd t}\Big)+p_\varphi \dd\Big({\dd \varphi
\over \dd t}\Big)\ ,\end{equation}
the associated Hamiltonian function will be given by an expression of
the standard form
\begin{equation} H= p_\varphi{\dd \varphi\over \dd t} + 
p_r{\dd r\over\dd t} -L \, .\end{equation}
It is to be remarked that this ordinary kind of Hamiltonian, as
defined with respect to the background time coordinate $t$, is to
be distinguished from the alternative kind of Hamiltonian defined 
with respect to the proper time $\tau$ (as used in the Wheeler 
De Witt type formulation~\cite{ANO00}), whose use requires the 
imposition of an extra constraint (whereby the proper time Hamiltonian 
is initially set to zero in the standard version of the Wheeler De Witt 
formulation).  In the present case the relevant momenta are
found to be given by the formulae
\begin{equation} \label{mom} p_\varphi = \overline n\, r^3
\, ,\qquad p_r=\left({\cal V} -{\cal V}^{-1}
\Big({\dd r\over \dd t}\Big)^2\right)^{-1/2}
 {\cal V}^{-1/2}\, {\cal U}\, r^3{\dd r\over
\dd t}\, ,\end{equation}
in which ${\cal U}$ is the energy density function given
according to (\ref{76}) by
\begin{equation} {\cal U}=\tn\, .\end{equation}
We thus obtain the Hamiltonian in the form
\begin{equation} H=  
\left({\cal V} -{\cal V}^{-1}
\Big({\dd r\over \dd t}\Big)^2\right)^{-1/2}\,{\cal V}\,{\cal U} r^3+
e_{_{[4]}}\, F\,{r^4\over 4} \, .\end{equation}

The time independence of the background ensures that the motion will be
such that this Hamiltonian has a constant value
\begin{equation} H=E_{_{\rm B}}\, ,\end{equation} 
that is identifiable as the integration constant introduced in
(\ref{70}), which is thus interpretable as a cosmological energy
density. A more trivial observation is that since the Lagrangian does 
not depend on the undifferentiated scalar  $\varphi$ but only on its 
gradient (i.e. since (\ref{momd}) contains no 
$\partial L/\partial \varphi$ term)
the corresponding momentum $p_\varphi$ will also be conserved, 
a conclusion that can be seen from (\ref{mom}) to be interpretable as 
equivalent to conservation of baryons, whose number density varies 
inversely as the cosmological volume, i.e. $\bar n\propto r^{-3}$.

As a final remark concerning energy conservation, it is to be observed 
that the stationary character of the background specified by (\ref{50}) 
and (\ref{Asol}) with respect to the action generated by the Killing 
vector $k^\mu$ given by $k_\mu\, \dd x^\mu=-{\cal V}\, \dd t$ ensures 
that even when inhomogeneous perturbations are taken into account there
will still be an energy density current, $\overline E^\mu$ say,
given by
\begin{equation} 
\overline E{^\mu}= k^\lambda\Big
(\overline T_{\!\lambda}{^\mu}
+{e_{_{[4]}}\over 4!} A_{\lambda\nu\rho\sigma}
{\cal E}^{\mu\nu\rho\sigma}\Big)\, , 
\end{equation}
that will satisfy a surface flux conservation law of the
simple form
\begin{equation} \overline\nabla_{\!\mu}\overline E{^\mu}=0\, .
\end{equation}

\end{document}